\documentclass[a4paper,aps,prd,10pt,preprintnumbers,showpacs,twocolumn,superscriptaddress,nofootinbib,amsmath,amssymb]{revtex4-1}
\usepackage{graphicx}
\usepackage{cmap}
\usepackage[utf8]{inputenc}
\usepackage[T1]{fontenc}
\usepackage{xcolor}

\def\imo{i}

\def\K{{\cal K}}

\begin{document}
\title{Asymptotic decay and quasinormal frequencies of scalar and Dirac fields around dilaton-de Sitter black holes}
\author{Alexey Dubinsky}
\email{dubinsky@ukr.net}
\affiliation{University of Seville, 41009 Seville, Spain}
\author{Antonina Zinhailo}
\email{antonina.zinhailo@physics.slu.cz}
\affiliation{Research Centre for Theoretical Physics and Astrophysics, Institute of Physics, Silesian University in Opava, Bezručovo nám. 13, CZ-74601 Opava, Czech Republic}
\begin{abstract}
We study the decay of Dirac and massive scalar fields at asymptotically late times in the background of the charged asymptotically de Sitter dilatonic black holes. It is shown that the asymptotic decay is exponential and oscillatory for large and intermediate mass of the field, while for zero and small mass it is pure exponential without oscillations. This reflects the dominance of quasinormal modes of the empty de Sitter spacetime at asymptotically late times.  We also show that the earlier WKB calculation of the massive scalar field spectrum 
does not allow one to find the fundamental mode with reasonable accuracy. 
\end{abstract}
\maketitle
\section{Introduction}

{Quasinormal modes have been a subject of intense study in recent years, primarily because they are observed by gravitational interferometers \cite{LIGOScientific:2016aoc,LIGOScientific:2017vwq,LIGOScientific:2020zkf}, with promising observations across broader frequency ranges anticipated in the near future \cite{Babak:2017tow}. Combining observations in the gravitational spectrum with those in the electromagnetic spectrum \cite{EventHorizonTelescope:2019dse,Goddi:2016qax} offers the potential to test gravitational theory in the strong field limit.}

{In addition to these observational perspectives, we focus on a particular theory of interest: the Einstein-Maxwell theory coupled with a scalar (dilaton) field and a non-zero cosmological constant. This theory is compelling due to the correspondence between quantum gravity in (anti)-de Sitter spacetime and conformal field theory \cite{Witten:2001kn,Strominger:2001pn}. In this context, the quasinormal modes of black holes can be interpreted as the poles of the Green functions in the dual field theory \cite{Birmingham:2001pj}. This correspondence describes the relaxation of quantum fields at finite temperature \cite{Son:2007vk}, which coincides with the Hawking temperature at the black hole event horizon.}

Quasinormal modes of various dilaton-like black holes, including those coupled to the higher curvature corrections, were considered in numerous publications \cite{Ferrari:2000ep,Carson:2020ter,Malybayev:2021lfq,Pani:2009wy,Konoplya:2019hml,Zinhailo:2019rwd,Lopez-Ortega:2009jpx,Kokkotas:2017ymc,Konoplya:2001ji,Fernando:2003wc,Chen:2005rm,Lopez-Ortega:2005obq,Paul:2023eep,Blazquez-Salcedo:2020caw,Pierini:2022eim}.
At the same time, quasinormal modes of asymptotically de Sitter black holes, unlike asymptotically flat case, govern the evolution of perturbations not only during the intermediate stage, but also at asymptotically late times $t \rightarrow \infty$ \cite{Dyatlov:2011jd,Dyatlov:2010hq,Konoplya:2024ptj}. In the asymptotically flat case, the stage of quasinormal ringing is changed by the asymptotic power-law tails.

Quasinormal modes of a scalar field in the dilaton black hole with de Sitter asymptotic were studied in \cite{Konoplya:2022zav}, though with an emphasis {on} charged and massless scalar fields, that is, in the limit when the mass of the field is negligibly small. On the other hand, a massive term brings qualitatively new features to the quasinormal spectrum, such as long-lived quasinormal modes \cite{Ohashi:2004wr,Konoplya:2017tvu,Churilova:2019qph}, oscillatory asymptotic tails \cite{Zinhailo:2024jzt}, or superradiant instability \cite{Konoplya:2008hj}. Notice that perturbations of massless fields may acquire an effective mass term owing to the brane-world scenarios \cite{Seahra:2004fg,Ishihara:2008re} or the presence of an external magnetic field \cite{Kokkotas:2010zd}.

Here we complement the work \cite{Konoplya:2022zav} in three aspects: First, we calculate the quasinormal modes of the massless Dirac field. Then, we study in detail the frequencies of a massive scalar field, including the case of a large product of the field's mass $\mu$ and black hole mass $M$. Finally, via time-domain integration, we determine the decay of these fields at asymptotically late times $t \rightarrow \infty$. Notice that the quasinormal modes of a massive scalar field were considered for these black holes also in \cite{Fernando:2016ftj}, though the results obtained there are shown only for $\ell \geq 1$ and some fixed values of the other parameters.{ Moreover, }as we will show in our work, the 6th order WKB approach used there suffers from considerable inaccuracy for $\ell=0$ perturbations, which may be much greater than the effect itself, leading even to senseless results.

Our work is organized as follows. In sec. II we summarize the main information about the theory under consideration, the black hole metric and the wave-like equations for scalar and Dirac fields. Sec. III is devoted to description of the WKB and time-domain integration methods used in this work. Sec. IV discusses the obtained numerical data on quasinormal frequencies. Finally, in the Conclusion we summarize the obtained results and review open questions.

\section{The black hole metric and wave equations}\label{sec:wavelike}

\begin{figure}
\resizebox{\linewidth}{!}{\includegraphics{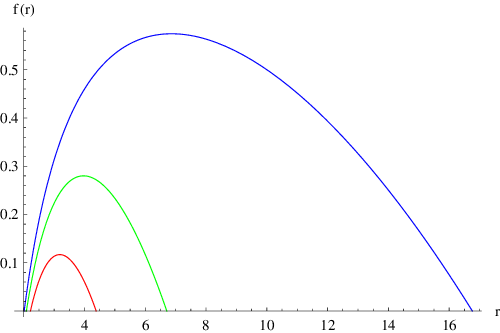}}
\caption{The dependence of the metric function $f(r)$ on the cosmological constant $\Lambda$ for fixed $Q=1/2$:  $\Lambda=1/100$ (blue), $\Lambda=55/1000$ (green), and $\Lambda=11/100$ (red).}\label{fig:lapselambda}
\end{figure}

\begin{figure}
\resizebox{\linewidth}{!}{\includegraphics{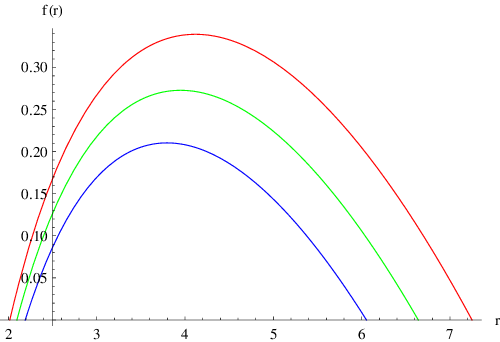}}
\caption{The dependence of the metric function $f(r)$ on the dilation charge $Q$ for fixed $\Lambda=55/1000$:  $Q=1/100$ (blue), $Q=45/100$ (green), and $Q=9/10$ (red).}\label{fig:lapseq}
\end{figure}

The first charged black hole solution in the dilaton gravity was obtained in \cite{Gibbons:1987ps} and later, independently in \cite{Garfinkle:1990qj}. This black hole is a solution to the low energy string theory, which is known now as the Gibbons-Maeda-Garfinkle-Horowitz-Strominger {(GMGHS)} black hole.
When the potential of the dilaton field is zero, no asymptotically de Sitter solution is allowed. 
In {a more} general case, the action for the dilaton gravity has the form,
\begin{equation}
S = \int d^4 x \sqrt{-g } \left[ R - 2 \partial_{\mu} \Phi \partial^{\mu} \Phi - V(\Phi) - e^{- 2 \Phi} F_{\mu \nu} F^{ \mu \nu} \right],
\end{equation}
where $R$ is the scalar curvature, $F_{\mu \nu}$ is the Maxwell's field strength and $\Phi$ is the dilaton field. 
The potential for the dilaton field is \cite{Gao:2004tu},
\begin{equation} \label{potential}
V(\Phi) = \frac{ 4 \Lambda}{3} + \frac{ \Lambda}{3} \left( e^{ 2 ( \Phi - \Phi_0)} + e^{ - 2 ( \Phi - \Phi_0)} \right).
\end{equation}
The exact solution describing  asymptotically de Sitter black hole in the dilaton theory was found in \cite{Gao:2004tu}
\begin{equation} \label{metric}
ds^2 = - f(r) dt^2 + \frac{ dr^2}{ f(r)} + R(r)^2 ( d \theta^2 + sin^2 \theta d \phi^2)
\end{equation}
where the metric functions are,
\begin{equation}
f(r) = 1 - \frac{ 2 M} { r} - \frac{ \Lambda r}{ 3} ( r - 2 Q), 
\end{equation}
\begin{equation}
R(r)^2 = r ( r - 2 Q).
\end{equation}
Here, $\Lambda$ is the cosmological constant, $M$ is the mass of the black hole and, $Q$ is the dilation charge.
When $\Lambda =0$, the metric given by eq. (\ref{metric}) reduces to the GMGHS black hole \cite{Gibbons:1987ps} \cite{Garfinkle:1990qj}.
When $Q=0$, the space-time {is reduced} the Schwarzschild-de Sitter solution. We will use units $M=1$.
{Examples of the dependence of the metric function $f(r)$ on $Q$ and $\Lambda$ are given in figs. \ref{fig:lapselambda} and \ref{fig:lapseq}, where one can see that the increased cosmological constant $\Lambda$ suppresses the metric function, but the charge  $Q$, on the contrary, enhances it.}

{For the above solution,} the dilation field $\Phi$, dilation charge $Q$, and electric field $F_{01}$, are given by the following relations,
\begin{equation}
e^{2 \Phi} = e^{ 2 \Phi_0} \left( 1 - \frac{ 2 Q}{r} \right),
\end{equation}
\begin{equation}
Q= \frac{ q^2 e^{ 2 \Phi_0} }{ 2 M}, \quad F_{01} = \frac{ q e^{ 2 \Phi_0}}{r^2},
\end{equation}
where
$\Phi_0$ is the dilation field at $r \rightarrow \infty$ and $q$ is the electric charge of the black hole.

The dynamical equations for a massive scalar $\phi$ and massless Dirac $\Upsilon$ fields in curved spacetime have the form:
\begin{subequations}\label{coveqs}
\begin{eqnarray}\label{KGg}
\frac{1}{\sqrt{-g}}\partial_\mu \left(\sqrt{-g}g^{\mu \nu}\partial_\nu\mathbb{\phi}\right)-\mu^2\mathbb{\phi}&=&0,
\\\label{covdirac}
\gamma^{\alpha} \left( \frac{\partial}{\partial x^{\alpha}} - \Gamma_{\alpha} \right) \Upsilon&=&0,
\end{eqnarray}
\end{subequations}
where $\gamma^{\alpha}$ are noncommutative gamma matrices, $\Gamma_{\alpha}$ are spin connections in the tetrad formalism and $\mu$ is the field's mass.
After separation of variables and introduction of a new wave function $\Psi$, the above equations (\ref{coveqs}) can be reduced to the following wavelike form \cite{Kokkotas:1999bd,Berti:2009kk,Konoplya:2011qq}:
\begin{equation}\label{wave-equation}
\dfrac{d^2 \Psi}{dr_*^2}+(\omega^2-V(r))\Psi=0,
\end{equation}
where the ``tortoise coordinate'' $r_*$ is:
\begin{equation}\label{tortoise}
dr_*\equiv\frac{dr}{f(r)}.
\end{equation}

The effective potentials for the scalar ($s=0$) field have the form
\begin{equation}\label{potentialScalar}
V(r)=f(r)\left(\frac{\ell(\ell+1)}{R(r)^2}+\mu^2\right)+\frac{1}{R(r)}\cdot\frac{d^2 R(r)}{dr_*^2},
\end{equation}
where $\ell=0, 1, 2, \ldots$ are the multipole numbers.
For the Dirac field one has two isospectral potentials,
\begin{equation}
V_{\pm}(r) = W^2\pm\frac{dW}{dr_*}, \quad W\equiv \left(\ell+\frac{1}{2}\right)\frac{\sqrt{f(r)}}{R(r)},
\end{equation}
where $\ell=1/2, 3/2, 5/2, \ldots.$
The isospectral wave functions {for the above two potentials $V_{\pm}$} can be transformed one into another by the well-known Darboux transformation,
\begin{equation}\label{psi}
\Psi_{+}\propto \left(W+\dfrac{d}{dr_*}\right) \Psi_{-}.
\end{equation}
Because of {this} isospectrality, we will study quasinormal modes for only one of the pair of the effective potentials, $V_{+}(r)$, because the WKB method is usually more accurate in this case.

\begin{figure}
\resizebox{\linewidth}{!}{\includegraphics{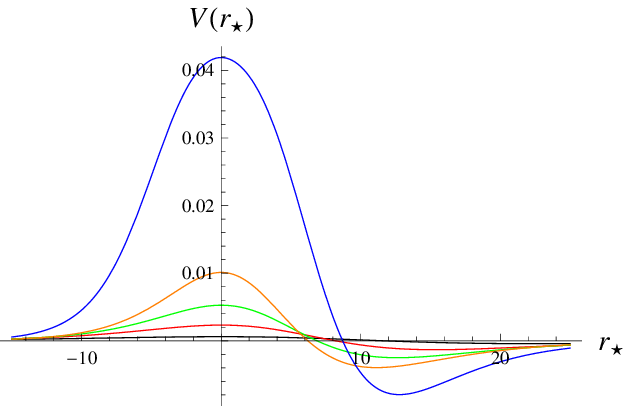}}
\caption{Effective potential of a scalar field as a function of the tortoise coordinate ($\ell=0$, $\mu=0$, $M=1$, $\Lambda=0.1$) for the dilaton-de Sitter black hole: $Q=0$ (black), $Q=0.2$ (red), $Q=0.4$ (green), $Q=0.6$ (orange), and $Q=0.99$ (blue).}\label{fig:potentials}
\end{figure}

\begin{figure}
\resizebox{\linewidth}{!}{\includegraphics{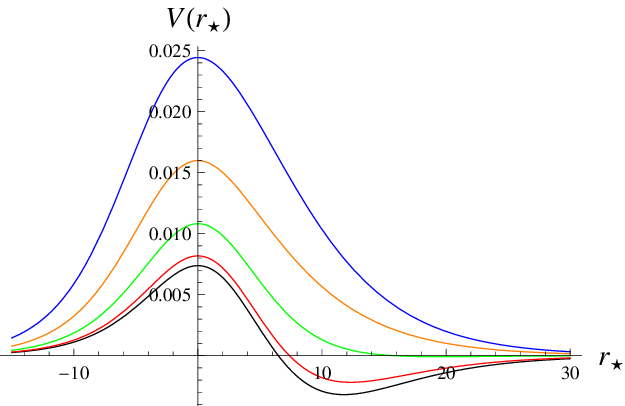}}
\caption{Effective potential of a scalar field as a function of the tortoise coordinate ($\ell=0$, $Q=0.5$, $M=1$, $\Lambda=0.1$)  for the dilaton-de Sitter black hole: $\mu=0$ (black), $\mu=0.1$ (red), $\mu=0.2$ (green), $\mu=0.3$ (orange), $\mu=0.4$ (blue).}\label{fig:potentials2}
\end{figure}

\begin{figure}
\resizebox{\linewidth}{!}{\includegraphics{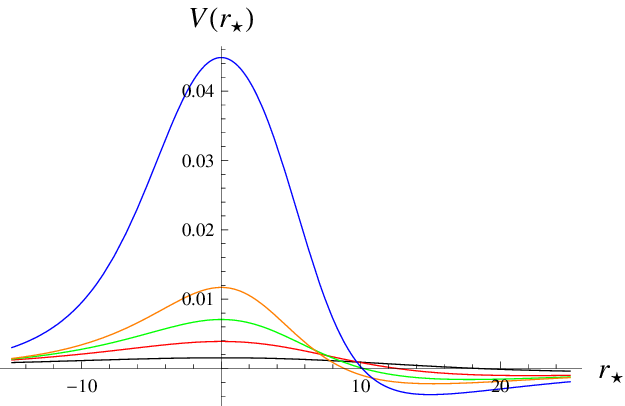}}
\caption{Effective potential  ($V_{+}$) of a Dirac field as a function of the tortoise coordinate ($\ell=1/2$, $\mu=0$, $M=1$, $\Lambda=0.1$) for the dilaton-de Sitter black hole: $Q=0$ (black), $Q=0.2$ (red), $Q=0.4$ (green), $Q=0.6$ (orange), and $Q=0.99$ (blue).}\label{fig:potentials3}
\end{figure}

\begin{figure}
\resizebox{\linewidth}{!}{\includegraphics{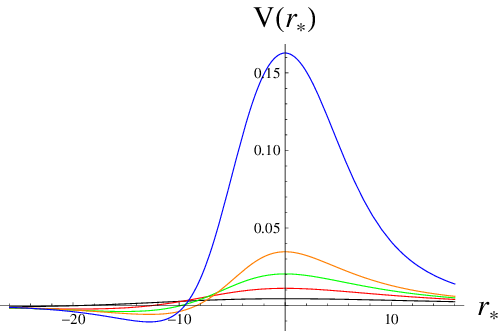}}
\caption{Effective potential  ($V_{-}$) of a Dirac field as a function of the tortoise coordinate ($\ell=1/2$, $\mu=0$, $M=1$, $\Lambda=0.1$) for the dilaton-de Sitter black hole: $Q=0$ (black), $Q=0.2$ (red), $Q=0.4$ (green), $Q=0.6$ (orange), and $Q=0.99$ (blue).}\label{fig:potentials4}
\end{figure}

\begin{figure}
\resizebox{\linewidth}{!}{\includegraphics{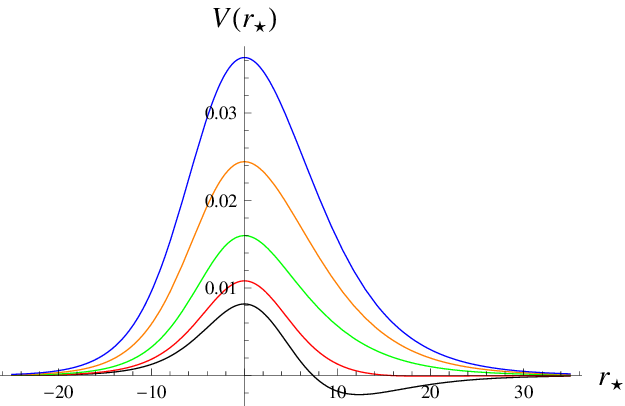}}
\caption{Potential as a function of the tortoise coordinate of the $\ell=0$ scalar field for the dilaton-de Sitter black hole ($M=1$, $Q=0.5$, $\Lambda=1/10$): $\mu=0$ (black), $\mu=0.5$ (red), $\mu=1$ (green), $\mu=3$ (orange), and $\mu=5$ (blue).}\label{fig:potentials5}
\end{figure}

\begin{figure}
\resizebox{\linewidth}{!}{\includegraphics{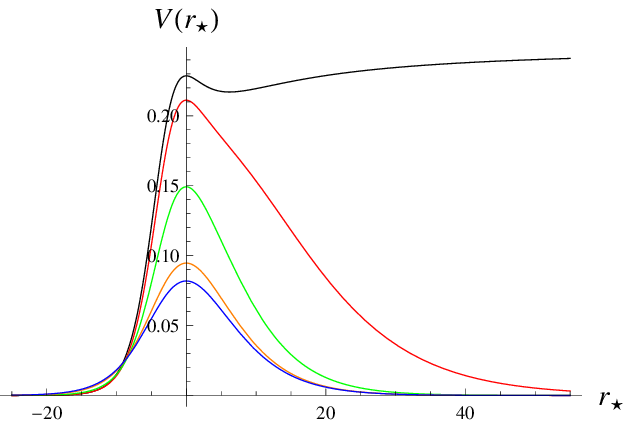}}
\caption{Potential as a function of the tortoise coordinate of the $\ell=1$ scalar field for the dilaton-de Sitter black hole ($M=1$, $\mu=0.5$, $Q=0.5$): $\Lambda=0$ (black) $\Lambda=0.01$ (red) $\Lambda=0.05$ (green) $\Lambda=0.09$ (orange), and $\Lambda=0.1$ (blue).}\label{fig:potentials6}
\end{figure}

{From figs. \ref{fig:potentials}-\ref{fig:potentials5} we see that larger $Q$ or $\mu$ correspond to a much higher peak of the effective potential. The dependence of the effective potential on the field's mass $\mu$ is totally different from the asymptotically flat case, for which the potential tends to a constant at infinity (see fig. \ref{fig:potentials6} and there is no maximum for sufficiently large $\mu$. 
Here we see that mass $\mu$ always provides a distinctive peak. However, for small $\mu$ there is also a negative gap at a distance from the black hole near the de Sitter horizon. Larger values of $\Lambda$ lead to a smaller height of the effective potential, as shown in fig. \ref{fig:potentials6}.}

\section{Methods used for calculation of quasinormal modes}\label{sec:WKB}

Quasinormal modes are complex frequencies $\omega$ which satisfy the following boundary conditions {for the wave function},
\begin{equation}\label{boundaryconditions}
\Psi(r_{*}\to\pm\infty)\propto e^{\pm\imo \omega r_{*}}.
\end{equation}
{These} are purely ingoing wave at the event horizon and purely outgoing waves at the de Sitter horizon.
{Here we will briefly discuss the two methods for analysis of quasinormal modes: WKB method and time-domain integration.}

\subsection{WKB Method}

The WKB method consists in matching of the asymptotic WKB solutions at both horizons with the Taylor expansion near the peak of the potential barrier. The first-order (eikonal) WKB formula is exact in the limit $\ell \to \infty$. The general WKB formula can be written as an expansion around the eikonal limit \cite{Konoplya:2019hlu}:
\begin{eqnarray}\label{WKBformula-spherical}
\omega^2&=&V_0+A_2(\K^2)+A_4(\K^2)+A_6(\K^2)+\ldots\\\nonumber&-&\imo \K\sqrt{-2V_2}\left(1+A_3(\K^2)+A_5(\K^2)+A_7(\K^2)\ldots\right).
\end{eqnarray}
The matching conditions for the quasinormal frequencies require that
\begin{equation}
\K=n+\frac{1}{2}, \quad n=0,1,2,\ldots.
\end{equation}
Here $n$ is the overtone number, $V_0$ is the maximal value of the effective potential, $V_2$ is the value of the second derivative of the potential in the maximum respectively the tortoise coordinate, $A_i$ for $i=2, 3, 4, \ldots$ is $i-th$ WKB order correction term beyond the eikonal approximation which depends on $\K$ and derivatives of the potential up to the order $2i$. 

{At intermediate and large $\mu M$  and non-zero $\Lambda$} the effective potential $V(r)$ of the master wave equation is  {a positive definite} barrier with a single peak, {as shown in fig. \ref{fig:potentials6}.}  Therefore, in this regime the WKB  approach can be applied for estimating the dominant modes. {However, for sufficiently small $\mu M$, there is a negative gap, so that the WKB method is much less accurate in this regime.}

The explicit form of the correction terms $A_i$ are deduced in \cite{Iyer:1986np} for the second and third WKB orders, in \cite{Konoplya:2003ii} for the 4-6th WKB orders and in \cite{Matyjasek:2017psv} for the 7-13th orders. 
Here we use the also further improvement of the accuracy of the above WKB formula via applying Padé approximants \cite{Matyjasek:2017psv} to the WKB series expansion. Following  \cite{Matyjasek:2017psv} we define a polynomial $P_k(\epsilon)$ via introducing powers of the order parameter $\epsilon$ in the righthand side of the WKB formula (\ref{WKBformula-spherical}):
\begin{eqnarray}\nonumber
  P_k(\epsilon)&=&V_0+A_2(\K^2)\epsilon^2+A_4(\K^2)\epsilon^4+A_6(\K^2)\epsilon^6+\ldots\\&-&\imo \K\sqrt{-2V_2}\left(\epsilon+A_3(\K^2)\epsilon^3+A_5(\K^2)\epsilon^5\ldots\right).\label{WKBpoly}
\end{eqnarray}
Here $k$ is the order of the WKB expansion. The parameter $\epsilon$ is defined as in \cite{Iyer:1986np} for
tracking of orders of the WKB series, and $\omega^2$ is found by taking $\epsilon=1$,
$$\omega^2=P_k(1).$$

We consider a set of  Padé approximants $P_{\tilde{n}/\tilde{m}}(\epsilon)$ for the polynomial $P_k(\epsilon)$ near $\epsilon=0$ with $\tilde{n}+\tilde{m}=k$ via constructing rational functions
\begin{equation}\label{WKBPadé}
P_{\tilde{n}/\tilde{m}}(\epsilon)=\frac{Q_0+Q_1\epsilon+\ldots+Q_{\tilde{n}}\epsilon^{\tilde{n}}}{R_0+R_1\epsilon+\ldots+R_{\tilde{m}}\epsilon^{\tilde{m}}},
\end{equation}
such that
$$P_{\tilde{n}/\tilde{m}}(\epsilon)-P_k(\epsilon)={\cal O}\left(\epsilon^{k+1}\right).$$
The rational function $P_{\tilde{n}/\tilde{m}}(\epsilon)$ is used then in order to approximate $\omega^2$,
\begin{equation}\label{omegaPadé}
\omega^2=P_{\tilde{n}/\tilde{m}}(1).
\end{equation}
Here we will use $\tilde{m} =4$, because such a splitting leads to the best accuracy for the Schwarzschild case. 
Overall, the WKB approach at various orders, with and without Padé approximants was used in numerous papers,
(see, for instance, \cite{Konoplya:2005sy,Konoplya:2006ar,Konoplya:2018ala,Chen:2023akf,Fu:2022cul,del-Corral:2022kbk,Barrau:2019swg,Guo:2023ivz}). Therefore we will not discuss it in more detail.

\subsection{Time-domain integration}
When $\mu M$ is small, we will use the time-domain integration of the wave equation according to the Gundlach-Price-Pullin discretization scheme \cite{Gundlach:1993tp}
\begin{eqnarray}
\Psi\left(N\right)&=&\Psi\left(W\right)+\Psi\left(E\right)-\Psi\left(S\right)\nonumber\\
&&- \Delta^2V\left(S\right)\frac{\Psi\left(W\right)+\Psi\left(E\right)}{4}+{\cal O}\left(\Delta^4\right),\label{Discretization}
\end{eqnarray}
Here, the points are defined as follows: $N\equiv\left(u+\Delta,v+\Delta\right)$, $W\equiv\left(u+\Delta,v\right)$, $E\equiv\left(u,v+\Delta\right)$, and $S\equiv\left(u,v\right)$. This method also was used in a lot of works \cite{Dubinsky:2024jqi,Konoplya:2005et,Konoplya:2018yrp,Varghese:2011ku,Abdalla:2012si,Momennia:2022tug} and is known to provide good accuracy for the fundamental mode after usage the Prony method for extracting the frequencies from the time-domain profile. 
Extraction of frequencies belonging to the Schwarzschild branch for $\ell=0$ is slightly less accurate, but still robust, because there are only a few oscillations which quickly go over into the asymptotic decay governed by the de Sitter branch of modes.

\section{Quasinormal modes and asymptotic fall-off}

\begin{figure}
\resizebox{\linewidth}{!}{\includegraphics{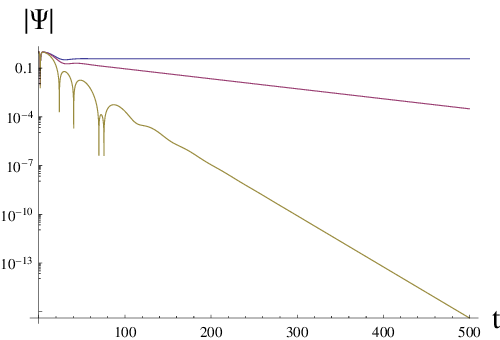}}
\caption{Time-domain profile for the scalar field ($\ell=0$) in the background of the dilaton-de Sitteer black hole $\mu M =0 $ (blue), $\omega_{Sch}=0.153303 - 0.107382 i$ (fitted within the region $t=25-50$), $\omega_{dS}=0 - 3.10464*10^{-8} i$ (fitted within $t=450-500$); $\mu M =0.1$ (magenta), $\omega_{Sch}=0.154844 - 0.106467 i$ ($t=25-50$), $\omega_{dS}=0 - 0.0141758 i$ ($t=450-500$); $\mu M =0.2$ (khaki), $\omega_{Sch}=0.156889 - 0.101298 i$ ($t=25-50$), $\omega_{dS}=0 - 0.0724723 i$ ($t=450-500$); $\Lambda M^{2} =0.1$, $Q/M = 0.9$.}\label{fig:timedomain0}
\end{figure}

\begin{figure}
\resizebox{\linewidth}{!}{\includegraphics{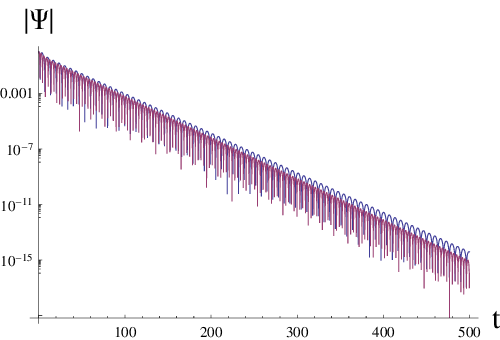}}
\caption{Time-domain profile for the scalar field ($\ell=0$) in the background of the dilaton-de Sitter black hole $\mu M =1 $ (blue), $\omega_{dS}=0.475065 - 0.0657918 i$ ($t=25-50$), $\omega_{dS}=0.475064 - 0.0657379 i$ ( $t=450-500$); $\mu M =2$ (magenta), $\omega_{dS}=0.957688 - 0.0683368 i$ ($t=25-50$), $\omega_{dS}=0 - 0.957651 - 0.0683422 i$ ($t=450-500$); $\Lambda M^{2} =0.1$, $Q/M = 0.9$.}\label{fig:timedomain1_2}
\end{figure}

\begin{figure}
\resizebox{\linewidth}{!}{\includegraphics{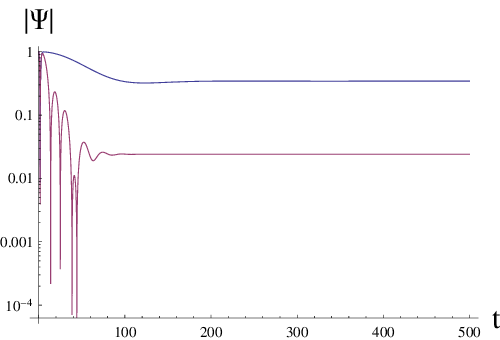}}
\caption{Time-domain profile for the Dirac field ($\ell=1/2$) in the background of the dilaton-de Sitter black hole $Q/M = 0$ (blue), $\omega_{Sch}=0.0528344 - 0.0252943 i$ ($t=30-50$), $\omega_{dS}=0 - 8.92496*10^{-8} i$ ($t=450-500$); $Q/M = 0.9$ (magenta), $\omega_{Sch}=0.284964 - 0.0850329  i$ ($t=25-50$), $\omega_{dS}=0 - 1.20847*10^{-8} i$ ($t=450-500$); $\Lambda M^{2} =0.1$. The fitting of the Schwarzschild mode at asymptotic times, unexpectedly, gives much better agreement with the WKB data: $\omega_{Sch}=0.285498 - 0.088472 i$ ($t=460-500$), while the WKB data is $\omega_{Sch}= 0.285479 - 0.088507 i$.}\label{fig:timedomaindirac}
\end{figure}

\begin{figure}
\resizebox{\linewidth}{!}{\includegraphics{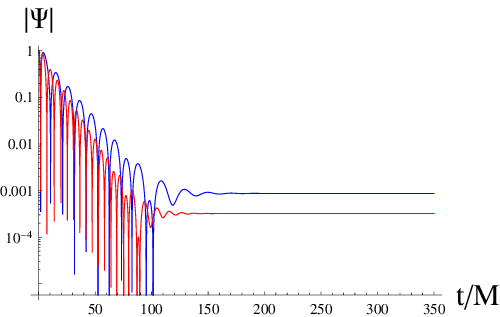}}
\caption{Time-domain profile for the Dirac field ($\ell=3/2$),$\Lambda M^{2} =0.1$ in the background of the dilaton-de Sitter black hole $Q/M = 1/2$ (blue), $\omega_{Sch}=0.304662 - 0.065726 i$ ($t=250-300$), $\omega_{dS}=0 - 1.59806*10^{-9} i$ ($t=250-300$); $Q/M = 0.9$ (red), $\omega_{Sch}=0.57349 - 0.08808 i$ ($t=250-300$), $\omega_{dS}=0 - 2.9284*10^{-9} i$ ($t=250-300$).}\label{fig:timedomaindirac32}
\end{figure}

\begin{table}
\begin{tabular}{c c c c c}
\hline
\hline
$Q$ & $\mu$ & WKB Padé ($\tilde{m}=4$) & WKB & difference \\
\hline
$0$ & $0$ & $0.019514-0.053922 i$ & $0.012614-0.049625 i$ & $14.2\%$\\
$0$ & $0.1$ & $0.045395-0.045689 i$ & $0.017504-0.044422 i$ & $43.3\%$\\
$0$ & $0.3$ & $0.047939-0.030064 i$ & $0.047435-0.031491 i$ & $2.68\%$\\
$0$ & $0.5$ & $0.088533-0.029366 i$ & $0.088562-0.029265 i$ & $0.113\%$\\
$0$ & $1.$ & $0.183544-0.029327 i$ & $0.183544-0.029327 i$ & $0.\times 10^{\text{-4}}\%$\\
$0$ & $2.$ & $0.370416-0.029358 i$ & $0.370416-0.029358 i$ & $0\%$\\
$0$ & $5.$ & $0.928400-0.029370 i$ & $0.928400-0.029370 i$ & $0\%$\\
$0$ & $10.$ & $1.857478-0.029372 i$ & $1.857478-0.029372 i$ & $0\%$\\
$0.5$ & $0$ & $0.064607-0.095158 i$ & $0.063479-0.093213 i$ & $1.96\%$\\
$0.5$ & $0.1$ & $0.068821-0.092728 i$ & $0.067374-0.089718 i$ & $2.89\%$\\
$0.5$ & $0.3$ & $0.107745-0.062028 i$ & $0.120141-0.051201 i$ & $13.2\%$\\
$0.5$ & $0.5$ & $0.183412-0.055714 i$ & $0.192229-0.047621 i$ & $6.24\%$\\
$0.5$ & $1.$ & $0.372149-0.055514 i$ & $0.372148-0.055523 i$ & $0.00248\%$\\
$0.5$ & $2.$ & $0.749040-0.056061 i$ & $0.749040-0.056061 i$ & $0.00002\%$\\
$0.5$ & $5.$ & $1.876499-0.056237 i$ & $1.876499-0.056237 i$ & $0\%$\\
$0.5$ & $10.$ & $3.754144-0.056262 i$ & $3.754144-0.056262 i$ & $0\%$\\
$0.9$ & $0$ & $0.145407-0.113036 i$ & $0.172296-0.068002 i$ & $28.5\%$\\
$0.9$ & $0.1$ & $0.146091-0.110539 i$ & $0.183920-0.059626 i$ & $34.6\%$\\
$0.9$ & $0.3$ & $0.163822-0.080109 i$ & $0.261706-0.013532 i$ & $64.9\%$\\
$0.9$ & $0.5$ & $0.245290-0.061271 i$ & $0.547789+0.015512 i$ & $123.\%$\\
$0.9$ & $1.$ & $0.475145-0.065984 i$ & $0.475089-0.066045 i$ & $0.0174\%$\\
$0.9$ & $2.$ & $0.957645-0.068343 i$ & $0.957645-0.068343 i$ & $0.00004\%$\\
$0.9$ & $5.$ & $2.399708-0.068880 i$ & $2.399708-0.068880 i$ & $0\%$\\
$0.9$ & $10.$ & $4.801038-0.068953 i$ & $4.801038-0.068953 i$ & $0\%$\\
\hline
\hline
\end{tabular}
\caption{Quasinormal modes of the $\ell=0$, $n=0$ scalar field for the dilaton-de Sitter black hole calculated using the 6th WKB order with and without Padé approximants; $M=1$, $\Lambda=0.1$.}
\end{table}

\begin{table}
\begin{tabular}{c c c c c}
\hline
\hline
$Q$ & $\mu$ & WKB Padé($\tilde{m}=4$) & WKB & difference \\
\hline
$0$ & $0$ & $0.081563-0.031217 i$ & $0.081560-0.031206 i$ & $0.0129\%$\\
$0$ & $0.1$ & $0.083621-0.031058 i$ & $0.083620-0.031041 i$ & $0.0188\%$\\
$0$ & $0.5$ & $0.123126-0.029676 i$ & $0.123127-0.029670 i$ & $0.0049\%$\\
$0$ & $1.$ & $0.202294-0.029335 i$ & $0.202294-0.029335 i$ & $0.\times 10^{\text{-4}}\%$\\
$0$ & $2.$ & $0.379979-0.029341 i$ & $0.379979-0.029341 i$ & $0\%$\\
$0$ & $5.$ & $0.932245-0.029366 i$ & $0.932245-0.029366 i$ & $0\%$\\
$0$ & $10.$ & $1.859401-0.029371 i$ & $1.859401-0.029371 i$ & $0\%$\\
$0.5$ & $0$ & $0.217683-0.069879 i$ & $0.217694-0.069906 i$ & $0.0129\%$\\
$0.5$ & $0.1$ & $0.220201-0.069213 i$ & $0.220203-0.069242 i$ & $0.0127\%$\\
$0.5$ & $0.5$ & $0.279305-0.059642 i$ & $0.279169-0.059882 i$ & $0.0968\%$\\
$0.5$ & $1.$ & $0.423675-0.055180 i$ & $0.423676-0.055181 i$ & $0.00015\%$\\
$0.5$ & $2.$ & $0.774382-0.055583 i$ & $0.774382-0.055584 i$ & $0.00002\%$\\
$0.5$ & $5.$ & $1.886536-0.056143 i$ & $1.886536-0.056143 i$ & $0\%$\\
$0.5$ & $10.$ & $3.759155-0.056238 i$ & $3.759155-0.056238 i$ & $0\%$\\
$0.9$ & $0$ & $0.426725-0.091004 i$ & $0.426744-0.090922 i$ & $0.0193\%$\\
$0.9$ & $0.1$ & $0.427989-0.090552 i$ & $0.428010-0.090470 i$ & $0.0194\%$\\
$0.9$ & $0.5$ & $0.460463-0.078966 i$ & $0.460471-0.078826 i$ & $0.0301\%$\\
$0.9$ & $1.$ & $0.585368-0.058865 i$ & $0.588662-0.056667 i$ & $0.673\%$\\
$0.9$ & $2.$ & $1.002604-0.064608 i$ & $1.002604-0.064608 i$ & $0.00001\%$\\
$0.9$ & $5.$ & $2.416892-0.068324 i$ & $2.416892-0.068324 i$ & $0\%$\\
$0.9$ & $10.$ & $4.809580-0.068816 i$ & $4.809580-0.068816 i$ & $0\%$\\
\hline
\hline
\end{tabular}
\caption{Quasinormal modes of the $\ell=1$, $n=0$ scalar field for the dilaton-de Sitter black hole calculated using the 6th WKB order with and without Padé approximants; $M=1$, $\Lambda=0.1$.}
\end{table}

\begin{table}
\begin{tabular}{c c c c c}
\hline
\hline
$Q$ & $\Lambda$ & WKB6 Padé & WKB6 & difference \\
\hline
$0$ & $0.01$ & $0.742849-0.036787 i$ & $0.742849-0.036787 i$ & $0.00003\%$\\
$0$ & $0.05$ & $0.481421-0.053523 i$ & $0.481426-0.053528 i$ & $0.00161\%$\\
$0$ & $0.1$ & $0.183544-0.029327 i$ & $0.183544-0.029327 i$ & $0.\times 10^{\text{-4}}\%$\\
$0$ & $0.11$ & $0.057035-0.009589 i$ & $0.057035-0.009589 i$ & $0\%$\\
$0.5$ & $0.01$ & $0.757890-0.036543 i$ & $0.757893-0.036541 i$ & $0.00047\%$\\
$0.5$ & $0.05$ & $0.546194-0.057499 i$ & $0.546208-0.057480 i$ & $0.00427\%$\\
$0.5$ & $0.1$ & $0.372149-0.055514 i$ & $0.372148-0.055523 i$ & $0.00248\%$\\
$0.5$ & $0.11$ & $0.338433-0.053025 i$ & $0.338428-0.053040 i$ & $0.00455\%$\\
$0.9$ & $0.01$ & $0.769909-0.036369 i$ & $0.769909-0.036366 i$ & $0.00037\%$\\
$0.9$ & $0.05$ & $0.593836-0.059915 i$ & $0.593842-0.059925 i$ & $0.00194\%$\\
$0.9$ & $0.1$ & $0.475145-0.065984 i$ & $0.475089-0.066045 i$ & $0.0174\%$\\
$0.9$ & $0.11$ & $0.455766-0.066100 i$ & $0.455656-0.066175 i$ & $0.0290\%$\\
\hline
\hline
\end{tabular}
\caption{Quasinormal modes of the $\ell=0$, $n=0$ scalar field perturbations for the dilaton-de Sitter black hole calculated using the 6th WKB order with and without Padé approximants; $M=1$, $\mu=1$.}
\end{table}

\begin{table}
\begin{tabular}{c c c c c}
\hline
\hline
$Q$ & $\Lambda$ & WKB Padé($\tilde{m}=4$) & WKB & difference \\
\hline
$0$ & $0$ & $0.182643-0.096566 i$ & $0.182646-0.094935 i$ & $0.790\%$\\
$0$ & $0.01$ & $0.175013-0.091801 i$ & $0.175034-0.090282 i$ & $0.769\%$\\
$0$ & $0.05$ & $0.138805-0.070803 i$ & $0.138871-0.069729 i$ & $0.690\%$\\
$0$ & $0.1$ & $0.060612-0.030366 i$ & $0.060624-0.030245 i$ & $0.180\%$\\
$0$ & $0.11$ & $0.019236-0.009623 i$ & $0.019236-0.009617 i$ & $0.026\%$\\
$0.5$ & $0$ & $0.229785-0.103221 i$ & $0.229418-0.102635 i$ & $0.274\%$\\
$0.5$ & $0.01$ & $0.223329-0.099928 i$ & $0.223034-0.099339 i$ & $0.270\%$\\
$0.5$ & $0.05$ & $0.194960-0.085863 i$ & $0.194919-0.085373 i$ & $0.231\%$\\
$0.5$ & $0.1$ & $0.150742-0.065376 i$ & $0.150743-0.065010 i$ & $0.223\%$\\
$0.5$ & $0.11$ & $0.139900-0.060563 i$ & $0.139909-0.060230 i$ & $0.219\%$\\
$0.9$ & $0$ & $0.337624-0.105976 i$ & $0.337897-0.105223 i$ & $0.226\%$\\
$0.9$ & $0.01$ & $0.332786-0.104322 i$ & $0.333013-0.103628 i$ & $0.209\%$\\
$0.9$ & $0.05$ & $0.312678-0.097517 i$ & $0.312758-0.097026 i$ & $0.152\%$\\
$0.9$ & $0.1$ & $0.285479-0.088507 i$ & $0.285475-0.088204 i$ & $0.101\%$\\
$0.9$ & $0.11$ & $0.279709-0.086626 i$ & $0.279698-0.086351 i$ & $0.0940\%$\\
\hline
\hline
\end{tabular}
\caption{Quasinormal modes of the $\ell=1/2$, $n=0$ Dirac field for the dilaton-de Sitter black hole calculated using the 6th WKB order with and without Padé approximants; $M=1$, $\mu=0$.}
\end{table}

\begin{table}
\begin{tabular}{c c c c c}
\hline
$Q$ & $\mu$ & WKB6 Padé & WKB6 & difference \\
\hline
$0$ & $0$ & $0.380041-0.096408 i$ & $0.380068-0.096366 i$ & $0.0128\%$\\
$0$ & $0.01$ & $0.362939-0.091878 i$ & $0.362961-0.091845 i$ & $0.0107\%$\\
$0$ & $0.05$ & $0.283463-0.071260 i$ & $0.283466-0.071250 i$ & $0.00353\%$\\
$0$ & $0.1$ & $0.121566-0.030408 i$ & $0.121566-0.030409 i$ & $0.0005\%$\\
$0$ & $0.11$ & $0.038485-0.009622 i$ & $0.038485-0.009622 i$ & $0\%$\\
$0.5$ & $0$ & $0.472210-0.102688 i$ & $0.472228-0.102636 i$ & $0.0113\%$\\
$0.5$ & $0.01$ & $0.458341-0.099570 i$ & $0.458357-0.099524 i$ & $0.0106\%$\\
$0.5$ & $0.05$ & $0.397767-0.086098 i$ & $0.397776-0.086069 i$ & $0.00729\%$\\
$0.5$ & $0.1$ & $0.304664-0.065726 i$ & $0.304666-0.065715 i$ & $0.00346\%$\\
$0.5$ & $0.11$ & $0.282241-0.060858 i$ & $0.282242-0.060851 i$ & $0.00272\%$\\
$0.9$ & $0$ & $0.679342-0.104642 i$ & $0.679341-0.104627 i$ & $0.00213\%$\\
$0.9$ & $0.01$ & $0.669517-0.103094 i$ & $0.669516-0.103080 i$ & $0.00213\%$\\
$0.9$ & $0.05$ & $0.628667-0.096683 i$ & $0.628667-0.096670 i$ & $0.00208\%$\\
$0.9$ & $0.1$ & $0.573487-0.088077 i$ & $0.573488-0.088066 i$ & $0.00192\%$\\
$0.9$ & $0.11$ & $0.561795-0.086260 i$ & $0.561796-0.086249 i$ & $0.00188\%$\\
\hline
\end{tabular}
\caption{Quasinormal modes of the $\ell=3/2$, $n=0$ Dirac field for
the dilaton-de Sitter black hole calculated using the 6th WKB formula
with and without Padé approximants; $M=1$, $\mu=0$.}
\end{table}

\begin{table*}
\begin{tabular}{c c c c c c}
\hline
\hline
$\mu $ & Prony fit & WKB Padé ($\tilde{m}=4$) & error &  WKB & error \\
\hline
$0.1$ & $-0.014176 i$ & $0.146091-0.110539 i$ & -- &  $0.183920-0.059626 i$ & --\\
$0.2$ & $-0.072457 i$ & $0.149710-0.100495 i$ & -- &  $0.222756-0.037417 i$ & --\\
$0.3$ & $0.167151-0.078907 i$ & $0.163822-0.080109 i$ & $1.91\%$ &  $0.261706-0.013532 i$ & $62.2\%$\\
$0.4$ & $0.203502-0.064919 i$ & $0.204251-0.067138 i$ & $1.10\%$ & $0.281788-0.049017 i$ & $37.4\%$\\
$0.5$ & $0.244383-0.060346 i$ & $0.245290-0.061271 i$ & $0.515\%$ &  $0.547789+0.015512 i$ & $124.\%$\\
$0.6$ & $0.287486-0.059885 i$ & $0.286833-0.058433 i$ & $0.542\%$ &  $0.253151-0.069894 i$ & $12.2\%$\\
$0.7$ & $0.332544-0.061215 i$ & $0.331233-0.061221 i$ & $0.388\%$ &  $0.327815-0.062493 i$ & $1.45\%$\\
$0.8$ & $0.379193-0.063017 i$ & $0.378607-0.063603 i$ & $0.216\%$ &  $0.377953-0.063905 i$ & $0.397\%$\\
$0.9$ & $0.426876-0.064593 i$ & $0.426786-0.065055 i$ & $0.109\%$ &  $0.426607-0.065184 i$ & $0.150\%$\\
$1.$ & $0.475066-0.065737 i$ & $0.475145-0.065984 i$ & $0.0540\%$ & $0.475089-0.066045 i$ & $0.0644\%$\\
\hline
\hline
\end{tabular}
\caption{Comparison of the quasinormal frequencies obtained by the Prony method from the time-domain profiles and those found by the 6th order WKB formula with and without Padé approximants for scalar field perturbations $\ell=0$, $Q=0.9$, $\Lambda M^2 =0.1$, $M=1$.}
\end{table*}

\begin{table*}
\begin{tabular}{c c c c c c}
\hline
\hline
$\mu $ & Prony fit & WKB Padé ($\tilde{m}=4$) & error &  WKB  & error \\
\hline
$0.1$ & $0.427994-0.090556 i$ & $0.427989-0.090552 i$ & $0.00126\%$ &  $0.428010-0.090470 i$ & $0.0198\%$\\
$0.2$ & $0.431820-0.089182 i$ & $0.431817-0.089181 i$ & $0.00058\%$ &  $0.431839-0.089099 i$ & $0.0192\%$\\
$0.3$ & $0.438322-0.086839 i$ & $0.438321-0.086840 i$ & $0.00038\%$ &  $0.438339-0.086760 i$ & $0.0181\%$\\
$0.4$ & $0.447738-0.083449 i$ & $0.447726-0.083452 i$ & $0.00285\%$ & $0.447734-0.083368 i$ & $0.0179\%$\\
$0.5$ & $0.460509-0.079004 i$ & $0.460463-0.078966 i$ & $0.0126\%$ &  $0.460471-0.078826 i$ & $0.0389\%$\\
$0.6$ & $0.477242-0.073783 i$ & $0.477237-0.073682 i$ & $0.0210\%$ &  $0.477446-0.073214 i$ & $0.125\%$\\
$0.7$ & $0.498368-0.068507 i$ & $0.498425-0.068410 i$ & $0.0224\%$ & $0.499611-0.067676 i$ & $0.297\%$\\
$0.8$ & $0.523773-0.064011 i$ & $0.523960-0.064062 i$ & $0.0367\%$ &  $0.523407-0.065422 i$ & $0.276\%$\\
$0.9$ & $0.552933-0.060785 i$ & $0.553016-0.060836 i$ & $0.0175\%$ &  $0.549478-0.062745 i$ & $0.714\%$\\
$1.$ & $0.585270-0.058903 i$ & $0.585368-0.058865 i$ & $0.0179\%$ &  $0.588662-0.056667 i$ & $0.691\%$\\
\hline
\hline
\end{tabular}
\caption{Comparison of the quasinormal frequencies obtained by the Prony method from the time-domain profiles and those found by the 6th order WKB formula withand without Padé approximants for scalar field perturbations $\ell=1$, $Q=0.9$, $\Lambda M^2 =0.1$, $M=1$.}
\end{table*}

\begin{table*}
\begin{tabular}{c c c c}
\hline
$\Lambda $ & Prony fit & WKB6 & error \\
\hline
$0$ & $0.230144-0.103140 i$ & $0.229785-0.103221 i$ & $0.146\%$\\
$0.01$ & $0.223675-0.099872 i$ & $0.223329-0.099928 i$ & $0.143\%$\\
$0.02$ & $0.216975-0.096529 i$ & $0.216636-0.096558 i$ & $0.143\%$\\
$0.03$ & $0.210021-0.093097 i$ & $0.209691-0.093095 i$ & $0.144\%$\\
$0.04$ & $0.202788-0.089569 i$ & $0.202476-0.089532 i$ & $0.142\%$\\
$0.05$ & $0.195240-0.085935 i$ & $0.194960-0.085863 i$ & $0.135\%$\\
$0.06$ & $0.187325-0.082182 i$ & $0.187098-0.082085 i$ & $0.121\%$\\
$0.07$ & $0.179004-0.078291 i$ & $0.178829-0.078180 i$ & $0.106\%$\\
$0.08$ & $0.170198-0.074230 i$ & $0.170077-0.074123 i$ & $0.0871\%$\\
$0.09$ & $0.160828-0.069961 i$ & $0.160752-0.069873 i$ & $0.0665\%$\\
$0.1$ & $0.150789-0.065433 i$ & $0.150742-0.065376 i$ & $0.0445\%$\\
$0.11$ & $0.139946-0.060561 i$ & $0.139900-0.060563 i$ & $0.0299\%$\\
\hline
\end{tabular}
\caption{Comparison of the time-domain fit and WKB formula for
$s=\ell=1/2$ ($M=1$, $Q=0.5$).}
\end{table*}

\begin{table*}
\begin{tabular}{c c c c}
\hline
$\Lambda $ & Prony fit & WKB6 & error \\
\hline
$0$ & $0.472212-0.102686 i$ & $0.472210-0.102688 i$ & $0.00046\%$\\
$0.01$ & $0.458342-0.099569 i$ & $0.458341-0.099570 i$ & $0.00046\%$\\
$0.02$ & $0.444014-0.096362 i$ & $0.444013-0.096363 i$ & $0.00041\%$\\
$0.03$ & $0.429179-0.093055 i$ & $0.429181-0.093057 i$ & $0.00069\%$\\
$0.04$ & $0.413784-0.089636 i$ & $0.413787-0.089640 i$ & $0.00117\%$\\
$0.05$ & $0.397763-0.086093 i$ & $0.397767-0.086098 i$ & $0.0015\%$\\
$0.06$ & $0.381034-0.082406 i$ & $0.381041-0.082413 i$ & $0.0024\%$\\
$0.07$ & $0.363495-0.078553 i$ & $0.363509-0.078564 i$ & $0.0049\%$\\
$0.08$ & $0.345039-0.074509 i$ & $0.345050-0.074524 i$ & $0.0054\%$\\
$0.09$ & $0.325490-0.070235 i$ & $0.325505-0.070260 i$ & $0.0088\%$\\
$0.1$ & $0.304617-0.065680 i$ & $0.304664-0.065726 i$ & $0.0210\%$\\
$0.11$ & $0.282207-0.060792 i$ & $0.282241-0.060858 i$ & $0.0257\%$\\
\hline
\end{tabular}
\caption{Comparison of the time-domain fit and WKB formula for
$s=1/2$, $\ell=3/2$ ($M=1$, $Q=0.5$).}
\end{table*}

The main feature of quasinormal spectrum of black holes in asymptotically de Sitter space is that there are essentially two branches of modes: perturbative in $\Lambda$, which tends to the Schwarzschild modes when $\Lambda \rightarrow 0$, and non-perturbative ones which tend to modes of empty de Sitter space when the black hole is small in comparison with the de Sitter radius. 

In the empty de Sitter spacetime, that is, when $M = 0$, the exact solution for quasinormal modes is known \cite{Lopez-Ortega:2012xvr, Lopez-Ortega:2007vlo}, which is given by the following two expressions:
\begin{equation}\label{exact1}
i \omega_{n} R = \ell + 2n + \frac{3}{2} \pm \sqrt{\frac{9}{4} - \mu^2 R^2},
\end{equation}
for
\begin{equation}
\frac{9}{4} > \mu^2 R^2,
\end{equation}
and
\begin{equation}\label{exact2}
i \omega_{n} R = \ell + 2n + \frac{3}{2} \pm i \sqrt{\mu^2 R^2-\frac{9}{4}},
\end{equation}
for
\begin{equation}
\frac{9}{4} < \mu^2 R^2.
\end{equation}
Here we used $R=\sqrt{3/\Lambda}$ is the de Sitter radius.

From the above equations (\ref{exact1}, \ref{exact2}) we see that at small $\mu$ the expression under the square 
root changes the sign, frequencies become purely imaginary. When the cosmological constant is small, that
is, the de Sitter radius is large and consequently the black hole is small relatively the de Sitter 
radius, the modes dominating at asymptotically late times are approaching those of the purely de Sitter space. 
That is why the frequencies in this regime are purely imaginary, as, for example, in table VI for $\mu = 0.1, 0.2$.

The dependence of the quasinormal modes upon the mass of the field $\mu$ is qualitatively different for small and large masses as well as depending whether they have non-zero oscillation frequency or not. 
For modes of small black holes with $Re (\omega) \neq 0$ the damping rate, given by the absolute value of $Im (\omega)$, decreases, while the real oscillation frequency increases (see tables I and II). When $\mu$ is large,  $Re (\omega)$ increases roughly proportionally 
to $\mu$, while the damping rate almost does not change. The latter happens,because the massive term dominates in the effective potential and, as could be shown via expansion in terms of $1/\mu$, similarly to eq. 22 in \cite{Konoplya:2024ptj}, the dominant term in the damping rate does not depend on $\mu$.  

From tables I - III we see that the charge $Q$ leads to the considerable increase of both real and imaginary parts of $\omega$. This happens for massive scalar and massless Dirac fields. {This increase can be easily explained by the behavior of the effective potentials, which are higher for larger values of $Q$.}

From time-domain profiles { (see figs. \ref{fig:timedomain0}-\ref{fig:timedomaindirac32}) } we see that no asymptotic tails take place at very late times $t/M \sim 500$, but  the quasinormal modes govern the decay, which is in concordance with observations made for gravitational and scalar perturbations of the four-dimensional Schwrazschild-de Sitter spacetime in \cite{Konoplya:2024ptj,Konoplya:2022xid}. {Indeed at late times we observe the domination of the purely de Sitter branch of modes both for the scalar (massless and massive) and massless Dirac perturbations, which are shown in examples of time-domain profiles  \ref{fig:timedomain0}-\ref{fig:timedomaindirac32}). The plots are semi-lograrithmic, so that the exponential decay at asymptotic time is clear and cannot be confused with the power-law tails which takes place in the asymptotically flat space.}
{Thus,} here we extend this conclusion about the dominance of quasinormal modes at asymptotic times also to the Dirac perturbations and to a charged dilatonic black hole.  

Another observation which we make is the poor accuracy of the usual WKB formula for $\ell=0$ scalar field perturbations. 
Comparison with the convergent time-domain integration method (see tables IV and V) shows that the usual WKB may have error exceeding a hundred percents. At the same time, the error of the 6th order WKB approach with $\tilde{m}=4$ Padé approximants is  $\approx 1.91 \%$ for the worst case $\mu =0.3$, and remains significantly below $1 \%$ for $\mu \gtrapprox 0.5$. Therefore, we can see that earlier publication by S. Fernando \cite{Fernando:2016ftj} which uses usual WKB formula of the 6th order, cannot include reliable analysis of even of the fundamental mode at the lowest multipole number $\ell=0$. It is worth mentioning that for $\ell=1$ we reproduce the results of table I in \cite{Fernando:2016ftj} when using the usual 6th order WKB formula.  
Notice that once the mass of the field is small or zero, the WKB method either usual, as in \cite{Fernando:2016ftj}, or with Padé approximants, cannot detect the lowest lying modes which dominate the fall-off at asymptotic times even for $\ell>0$, because the they have $Re (\omega) \ll \mid Im (\omega) \mid$   and the extension into the complex plain implies that the turning points are separated a lot and the Taylor expansion in the middle zone does not match well with the WKB solutions at the asymptotic regions \cite{Konoplya:2019hlu}. {Indeed, on figs.  \ref{fig:timedomain0}-\ref{fig:timedomaindirac32}) wee see such purely exponential mode dominating at asymptotic times which cannot be detected by the WKB method in principle. However, as shown in tables I-IX, quasinormal modes of the Schwarzschild branch can usually be well-approximated by the WKB method, unless the mass of the field is very small. For the massless Dirac field we can reproduce  the Schwarzschild branch of modes in tables VIII and IX, which are responsible for the first ringdown stage (see figs. \ref{fig:timedomaindirac} and \ref{fig:timedomaindirac32}).}

If the effective potential is positive definite in the whole range outside the black hole event horizon, the
differential operator
\begin{equation}
\Sigma = -\frac{\partial^{2}}{\partial r_{\star}^{2}} + V
\end{equation}
is a positive self-adjoint operator in the Hilbert space of square
integrable functions, and, consequently, all solutions of the perturbation equations with compact support initial
conditions must be bounded, which means the stability.

As can be seen from figs. \ref{fig:potentials}-\ref{fig:potentials5} the effective potentials even for the scalar and Dirac fields have a negative gap far from the black hole. Depicting them in terms of the tortoise coordinate ensures that the region between the event horizon and de Sitter horizon is under consideration. Thus, one cannot exclude that the bound state with negative energy, i.e. the instability, could exist. 
Indeed, instabilities of a scalar field are known in the presence of a black hole charge and non-zero cosmological constant \cite{Zhu:2014sya,Konoplya:2014lha}.

The time-domain integration run for various parameters does not show any growing profiles, what signifies the stability of the perturbation. The formally "unstable mode", i.e. with positive imaginary part,  obtained by the usual 6th order WKB method for $Q=0.9$, $\mu =0.5$, $\ell=0$ in table I is the result of a huge error, which is treated by application of the Padé approximants. Notice, that the WKB method does not allow to find true modes with a positive imaginary part in principle \cite{Konoplya:2019hlu}.

The cosmological constant $\Lambda$  suppresses both real and imaginary parts of the frequencies belonging to the Schwarzschild branch of the massless fields, as can be seen from table III. Similar behavior is observed for the Schwarzschild-de Sitter case \cite{Moss:2001ga,Zhidenko:2003wq,Konoplya:2004uk} {and it could be explained by the behavior of the effective potentials which become smaller for larger values of $\Lambda$.} It is also worth mentioning that while for asymptotically flat black holes the spectrum of massive fields of various spin contains arbitrarily long lived modes, called quasi-resonances \cite{Ohashi:2004wr,Konoplya:2006gq}, we do not observe such modes, similar, in a sense, to standing waves, when the cosmological constant is nonzero. 

\section{Conclusions}

We have calculated quasinormal modes of the massless Dirac and massive scalar fields in the background of the dilaton-de Sitter black hole. While to the best of our knowledge this is first calculation of the Dirac field's spectrum in the dilaton-de Sitter black hole background, the massive scalar field has been previously studied in  \cite{Fernando:2016ftj} with the usual 6th order WKB method.
Here, by comparison with the time-domain integration data, we showed that while the usual WKB formula used in \cite{Fernando:2016ftj} is relatively accurate for $\ell=1$ and higher, it cannot be applied to $\ell=0$ case, because then the error can achieve more than a hundred percents. Moreover, for small $\mu$ the purely imaginary frequencies dominate in the spectrum, which cannot be calculated by the above WKB method in principle.     
  
We complemented the previous publication on the scalar field spectrum  not only by studying $\ell=0$ scalar field pertubartions, but  also by consideration time-domain evolution of the perturbations. Particularly, we have shown that despite the effective potential has a negative gap, the perturbations are decayed, implying the stability of the field under consideration. In addition we show at asymptotic times $t \rightarrow \infty$ the de Sitter branch of quasinormal modes dominate in a signal. 

We have also studied quasinormal  frequencies of the massless Dirac field and shown that the asymptotic decay is again governed by the de Sitter branch of quasinormal modes. The fundamental mode is strongly suppressed when the cosmological constant is turned on. 

\begin{acknowledgments}
The authors are thankful to R. A. Konoplya for useful discussions. Alexey Dubinsky acknowledges the University of Seville for their support through the Plan-US of aid to Ukraine. Antonina  Zinhailo acknowledges the support from the Silesian University grant SGS/24/2024.
\end{acknowledgments}

\bibliographystyle{unsrt}
\bibliography{bibliography}
\end{document}